# Petrophysical analysis of regional-scale thermal properties for improved simulations of geothermal installations and basin-scale heat and fluid flow

Andreas Hartmann

*Applied Geophysics and Geothermal Energy, E.ON Energy Research Center, RWTH Aachen University, Lochnerstr. 4-20, D-52056 Aachen*

*Now at: Baker Hughes INTEQ, Baker-Hughes-Str. 1, 29221 Celle*

Tel: +49 5141 203 807

mailto:Andreas.Hartmann@inteq.com

http://www.bakerhughes.de

Renate Pechnig

*Geophysica Beratungsgesellschaft mbH, Lütticher Str. 32, D-52064 Aachen*

Tel: +49 241 189 929 15

Fax: +49 241 189 929 13

mailto:info@geophysica.de

http://www.geophysica.de

Christoph Clauser

*Applied Geophysics and Geothermal Energy, E.ON Energy Research Center, RWTH Aachen University, Lochnerstr. 4-20, D-52056 Aachen*

Tel: 49 241 809 4825

Fax +49 241 80 92132

mailto:c.clauser@geophysik.rwth-aachen.de

http://www.geophysik.rwth-aachen.de

**Abstract** Development of geothermal energy and basin-scale simulations of fluid and heat flow both suffer from uncertain physical rock properties at depth. For the production of geothermal energy, a high risk of failure is associated with this uncertainty. Invoking the usual conservative assumptions as a remedy results in unnecessarily large drilling depths and increased exploration costs. Therefore, building better prognostic models for geothermal installations in the planning stage requires improvement of this situation. To this end, we analysed systematically the hydraulic and thermal properties of the major rock types in the Molasse Basin in Southern Germany. On about 400 samples thermal conductivity, density, porosity, and sonic velocity were measured in the laboratory. The size of both the study area and the this data set require special attention with respect to the analysis and the reporting of data, in particular in view of making it useful and available for practitioners in the field. Here, we propose a three-step procedure with increasing complexity, accuracy, and insight into petrophysical relationships: First, univariate descriptive statistics provides a general understanding of the data structure, possibly still with large uncertainty. Examples show that the remaining uncertainty can be as high as 0.8 W $(m\ K)^{-1}$ or as low as 0.1 W $(m\ K)^{-1}$. This depends on the possibility to subdivide the geologic units into data sets that are also petrophysically similar. Then, based on all measurements, cross-plot and quick-look methods are used to gain more insight into petrophysical relationships and to refine the analysis. Because these measures usually imply an exactly determined system they do not provide strict error bounds. The final, most complex step comprises a full inversion of select subsets of the data comprising both laboratory and borehole measurements. The example presented shows the possibility to refine the used mixing laws for Petrophysical properties and the estimation of mineral properties. These can be estimated to an accuracy of 0.3 W $(m K)^{-1}$. The predictive errors for the measurements are 0.07 W $(m K)^{-1}$, 70 m $s^{-1}$, and 8 kg $m^{-3}$ for thermal conductivity, sonic velocity, and bulk density, respectively. The combination of these three approaches provides a comprehensive understanding of petrophysical properties and their interrelations, allowing to select an optimum approach with respect to both the desired data accuracy and the required effort.

# Introduction

Thermal and hydraulic properties of the subsurface play an important role in the modelling of heat and fluid transport, be it for the planning of geothermal installations or regional scale modelling of basins. In general, the ranges of these properties given in compilations of rock properties (e. g. Birch 1966; Čermák and Rybach 1982; Haenel et al. 1988; Clauser and Huenges 1995; Clauser 2006) are too wide to be useful to constrain properties at a specific site. To improve this situation, we performed a systematic study of hydraulic and thermal properties of major rock types in the Molasse Basin in Southern Germany combining a statistical approach based on a large number of laboratory measurements with the inversion of suitable combinations of geophysical borehole logs. When used in prognostic design calculations for new geothermal installations this new data base will help to reduce the risk of failure. In view of geodynamic applications, this data provides much improved constraints for basin-scale numerical simulations of heat and fluid flow.

To characterise the relevant petrophysical properties, we measured thermal and hydraulic properties on rock samples in the laboratory. In particular, thermal conductivity, heat capacity, porosity, density, and permeability where measured. In such an approach, the sample collection needs to be representative of the stratigraphic units studied. More often than not, this cannot be accomplished completely because core from boreholes is generally rare. Thus, additional data are required to ensure that statistical results are representative. Particularly useful data derive from wireline logging of hydrocarbon exploration wells. These geophysical downhole measurements can be used to complement the laboratory measurements. To this end, a specific petrophysical model is derived for a lithologic unit from the laboratory data. This model is then applied to the well log data to derive the desired properties. This way, not only the spatial coverage is increased but also the vertical lithologic profile is completed with data for a given geology.

The next section will describe the study area and its geologic units, followed by a summary of the used data and applied methods. Finally, examples of the petrophysical analysis of the data set are used to illustrate both the opportunities and the limitations of the aspired generalisation of petrophysical properties on a regional scale.

# Description of study area, data, and methods

The study is focused on the Western Molasse region in southern Germany and the Jurassic and Triassic landscapes north of the Molasse basin (Fig. 1). This region is particularly suited for a detailed due to its large technical potential for geothermal use (Erbas 1999) and high degree of geological exploration: The subsurface of the Molasse basin is well known due to the past and intense exploration for hydrocarbons. The northern part of the study area is characterized by a sequence of southward dipping Mesozoic rocks (Fig. 2). Thus, moving to the north, progressively older rocks can be sampled at the surface. Comparisons of Mesozoic samples taken from outcrops or shallow boreholes with those taken from the Molasse region enable to study PT-dependence and possible facies changes of the rocks.

**Fig. 1** Map of the study area (brown box) in southern Germany. Triangles and squares indicate locations for core sampling from boreholes and outcrops, respectively. Circles show locations of hydrocarbon wells.

**Fig. 2** Simplified geological cross section of the study area and the standard stratigraphic profiles of the two selected stratigraphic units Lower Triassic (Bunter) and Upper Jurassic (Malm) (after Geyer and Gwinner 1991).

This study comprises a total of about 400 core samples. About two thirds were obtained from core archives of the Geological Survey of Baden-Württemberg, the University of Tübingen, and Wintershall AG, Kassel. These samples were taken in boreholes for water or hydrocarbon exploration. The borehole samples where complemented by cores taken from outcrops. The sampled geologic sequence ranges from Lower Triassic (Bunter) to Miocene rocks. On all samples routine measurements were performed in the laboratory to determine thermal conductivity, density, porosity, and compressional wave velocity. The use of core scanning devices allowed rapid and accurate measurement of a large number of samples. Thermal conductivity was determined at ambient temperature using the optical scanning method. A detailed description of the technique can be found in Popov et al. (1999). The optical scanning lines were oriented parallel to the core vertical axis. Measurements were performed on dry and saturated rocks. The thermal conductivity along the core surface was recorded in 1 or 2 mm steps. An inhomogeneity factor $b = (\lambda_{\text{max}} - \lambda_{\text{min}}) / \lambda_{\text{avg}}$ was defined for each sample, computed from the maximum, minimum and average thermal conductivity of the scan line. Bulk density, grain density and porosity were determined from measurements of wet and dry sample mass and sample volume using Archimedes principle. By this way an average value for porosity, grain and bulk density is generated for each core sample. In addition gamma-density scans were performed by a multi sensor core logger (MSCL), providing bulk density values along the core axis. With the MSCL, bulk density is computed by measuring the attenuation of gamma rays that have passed through the cores, with the degree of attenuation being proportional to density (Boyce, 1976). Sonic velocity was measured on dry and saturated cores using the acoustic logging sensor of the MSCL. The device records the travel time of a 500 kHz pulse sent through the core. The sensors are aligned to measure the compressional wave speed perpendicular to the core axis. The sampling distance for all MSCL measurements ranged from 0.2 cm to 2 cm, the exact choice being a compromise between required resolution and available time.

Mineral content was derived by XRD-analysis performed on plugs (2 cm diameter) taken from of the core samples. This allowed more detailed analysis of petrophysical and mineralogical relationships. Prior to XRD analysis, grain density and porosity of the plug samples were derived with a helium pycnometer. For a detailed analysis of all data the reader is referred to Clauser et al. (2007).

The following analysis of petrophysical data will ignore the PT-dependences and focus mainly on two lithologies: Upper Jurassic (Malm) and Lower Triassic (Bunter). In addition, inversion methods will be discussed on a sample from the middle Triassic. The Upper Jurassic consists primarily of limestones with interspersed beds of marlstone, classified into sub-units alpha to zeta. Using gamma-ray logs, the sub-units Malm alpha to Malm delta can be correlated laterally. These show only minor variations in thickness and appearance. However, the Malm units zeta and epsilon show considerable variability due to changes between reef and massive facies. Also, the topmost subunits are eroded in boreholes in the north. The Lower Triassic (Bunter) formation of the study area is dominated by reddish sandstones deposited on a flat fluviatile plain under semi-arid conditions. Coarse grained deposits predominate in the southern branch of the Bunter basin in Germany. Clayey sediments only occur at the base of the Bunter and at the Röt formation, marking the marine transgression at the end of the Upper Bunter. The Bunter is subdivided into several sedimentary cycles. The grain-size variations are caused by the development of the fluviatile systems under changing relief energy. Thickness of the Bunter strongly changes within the study area. From the boundary of the south-eastern extension of the Bunter, the thickness rapidly increases up to 500 m in north-westerly direction.

# Analysis of petrophysical data

The large data set and the distribution of the sampling points over a large area make a sensible analysis challenging. Several methods were employed, all with specific advantages and disadvantages. The methods of analysis will be detailed in the following sections:

## Univariate descriptive statistics

The simplest method to analyse a large data set is to study its distribution. This yields mean values and uncertainties which can be used directly as input in design calculations for geothermal installations or in basin-scale simulations of heat and fluid flow. However, several aspects of this method may be undesirable: It must be ensured that the analysed sample is representative of the complete distribution. Further, the ensemble of samples needs to be both sufficiently large and uniform. These requirements are difficult to satisfy when considering a large area with variable geology.

**Table 1** Statistics of core scanning measurements on Lower Triassic samples. Compared are quartiles of thermal conductivity (in W (m K)$^{-1}$) for the Upper and Middle Bunter from two studies. Top row: Data measured in this study. Bottom row: Data from a previous field campaign (Kleiner 2003; Clauser et al. 2003).

As an example Table 1 shows the quartiles of thermal conductivity derived from measurements in this study. The top row contains all samples obtained from boreholes in the main study area (brown box in Fig. 1). The bottom row shows the statistics of a large data set from a previous study of the thermal properties of the Bunter near the town of Ettlingen in the Rhine Graben near Karlsruhe (Kleiner 2003; Clauser et al. 2003). Although both data sets sample the same stratigraphy, their median values differ considerably from each other, due to different facies types within the same stratigraphy. In this particular case it is well understood that different locations with different facies of one stratigraphy were sampled. In general it might be more difficult to decide whether a particular data set is truly representative of a larger region.

Assuming that the Bunter data set is representative, a mean value of 4.3 W (m K)$^{-1}$ ± 0.7 W (m K)$^{-1}$ results for the Lower Triassic. If this value was used in modelling assuming a $2\sigma$-probability, the corresponding 95 % confidence interval would be 2.9 W (m K)$^{-1}$ to 5.7 W (m K)$^{-1}$. In the lower Triassic, this large range results from the superposition of two effects: (1) the variable matrix conductivity due to the varying amounts of quartz, feldspars and micas within the Bunter sandstones; (2) the variable rock porosity of up to of 25 %. Obviously, it is desirable to narrow down this range.

**Fig. 3** Histogram of laboratory measurements of thermal conductivity for samples from the Upper Jurassic. Mean and standard deviation are 2.40 W (m K)$^{-1}$ ± 0.12 W (m K)$^{-1}$. Median, 25 % quartile, and 75 % quartile are 2.39 W (m K)$^{-1}$, 2.30 W (m K)$^{-1}$, and 2.48 W (m K)$^{-1}$, respectively. Total number of data points is 1582.

Dividing an entire formation into sub-units may help to increase the accuracy of the predicted thermal conductivity. Fig. 3 shows the histogram of thermal conductivity for the Upper Jurassic. The shape of the distribution is not truly Gaussian, nevertheless, using quartiles of the distribution, the 50 % confidence interval about the median is 2.30 W (m K)$^{-1}$ to 2.48 W (m K)$^{-1}$. By dividing the data set into stratigraphic sub-units the 50 % confidence interval (indicated by the blue boxes) can be narrowed down to roughly 0.1 W (m K)$^{-1}$ (Fig. 4).

It is interesting to note that the largest values of thermal conductivity are found for the Malm zeta formation with values between 2.6 W (m K)$^{-1}$ and 2.8 W (m K)$^{-1}$. The XRD analysis shows the samples to be a very clean limestone, consisting sometimes of up to 99 % calcite and with porosities as low as 4 %.  The thermal conductivity of single mineral calcite aggregates is tabulated as 3.6 W (m K)$^{-1}$ (Clauser 2006). Based on this value, the conductivity of a clean, low porosity limestone can estimated as 3.3 W (m K)$^{-1}$. This is 0.5 W (m K)$^{-1}$ larger than the measured values. Thus, estimating rock properties based on tabulated values may result in large errors. This emphasises the importance of characterizing thermal rock properties on a regional, statistically firm base.

**Fig. 4** Statistical variation of thermal conductivity for the three sub-units Gamma, Beta and Zeta of the Upper Jurassic. Blue boxes – quartiles; red lines – median, black whiskers – data range; red crosses – outliers. The data range encompasses all data points or a range of 1.5 times the distance between quartiles, whichever is smaller.  If points exist outside the data range, i.e. > 1.5 of the interquartile range, they are considered outliers.

## Quick look methods using petrophysical models

So far, our univariate analysis was based on thermal conductivity measurements only. Now, we include other standard measurements in order to derive thermal conductivity from petrophysical models. In contrast to the statistical method which is based directly on measured values, this enables to determine thermal conductivity from secondary data which may be readily available.

To utilise these other data, it is necessary to develop or adapt petrophysical models, to jointly describe the measured properties of the sample. Generally, a mixing law is used which weights the contributions of pore fluid and rock matrix by porosity $\phi$. Pore fluid and matrix properties are denoted by indices *f* and *m*, respectively. An often used empirical law for acoustic slowness $\Delta t$ is given by (Wyllie 1956):

$$\Delta t = \phi\, \Delta t_f + (1-\phi)\Delta t_m \,. \tag{1}$$

Density $\rho$ is described as:

$$\rho = \phi\, \rho_f + (1-\phi)\,\rho_m \,. \tag{2}$$

Finally, thermal conductivity $\lambda$ can be described by the empirical geometric mixing law (Woodside and Messmer, 1961):

$$\lambda = \lambda_f^{\phi}\, \lambda_m^{(1-\phi)} \,. \tag{3}$$

While this two-phase approach used allows for the rock matrix and the pore fluid, the solid rock phase itself may be composed of different mineral phases. The Upper Jurassic rocks, for instance, consist of calcite as the main phase and variable amounts of shale. Extending the mixing laws (1) and (3) for acoustic slowness and thermal conductivity, respectively, to this case yields:

$$\Delta t = \phi \Delta t_f + (1-\phi)\left(V_{SH}\Delta t_{SH} + (1-V_{SH})\Delta t_{LS}\right), \quad (4)$$

$$\lambda = \lambda_f^{\phi}\left(\lambda_{SH}^{V_{SH}} \cdot \lambda_{LS}^{(1-V_{SH})}\right)^{(1-\phi)}, \quad (5)$$

where subscripts LS and SH denote limestone (i.e. calcite) and shale, respectively, and $V_{SH}$ is the shale volume fraction of the solid phase. Fig. 5 shows a plot of thermal conductivity $\lambda$ versus slowness $\Delta t$ for the Upper Jurassic data set together with the theoretical values for the three end member water, shale, and calcite connected by the grey triangle. All measured data should plot within this triangle, and thus a rock's volumetric composition can be read directly from this plot. Vice versa, a rock's thermal conductivity can be inferred from this plot if its volumetric composition is known or can be estimated, for instance from its natural gamma activity which is sensitive to the shale volume.

**Fig. 5** Cross-plot of thermal conductivity (logarithmic scale) versus slowness (linear scale) for laboratory measurements from the Upper Jurassic. Colour coding corresponds to measured porosity. End member values for slowness are taken from Hearst et al. (2000), for thermal conductivity from Čermák and Rybach (1982).

In practice, values for the end member points cannot be simply adopted from numerical tables. Rather, the end member points need to be placed at reasonable positions as part of the interpretation. In particular, the shale point is generally poorly defined and needs to be adjusted to match the data. In the case shown in Fig. 5, shale slowness at 220 $\mu s\ m^{-1}$ is very low, and the limestone matrix thermal conductivity has a value of only 3.1 $W\ (m\ K)^{-1}$. But these choices for the end member points are suggested by the location of the plotted data.

The direct cross-plot of two measured properties is only possible in a system with three components, in this case calcite, shale, and pore fluid. If the number of components is larger than that, more measurements need to be taken into account and the analysis method needs to be changed. Burke et al. (1969) developed the M-N plot, a method to analyse the mineral composition of a three-phase system. The original method uses sonic, neutron porosity, and density logs to compute two parameters M and N which are independent of porosity and can be used to identify the occurring minerals:

$$M = \frac{\Delta t_f - \Delta t}{\rho - \rho_f}, \quad (6)$$

$$N = \frac{1-\phi}{\rho - \rho_f}, \quad (7)$$

where the subscript $f$ denotes fluid properties. Inserting equations (1) and (2) for $\Delta t$ and $\rho$, respectively, removes porosity from equations (6) and (7). Thus, a cross-plot of these two parameters is determined only by the matrix composition. A neutron porosity measurement is not possible for the samples, but the method can be adapted. We define a parameter O that uses the logarithm of the thermal conductivity in the same manner as the neutron porosity does:

$$O = \frac{\log_{10}\lambda_f - \log_{10}\lambda}{\rho - \rho_f}. \quad (8)$$

Taking the logarithm of the thermal conductivity measurement ensures that the value scales linearly with porosity in the same manner as density and slowness do. As an example, the Lower Triassic data set is analysed using this method. In Fig. 6, O is plotted versus M together with expected (M,O)-pairs. In addition to the (M,O) values derived from the petrophysical measurements, the mineral compositions from XRD-analysis on a sub-set of samples are shown in Fig. 6 as circles. The general appearance is satisfactory, but it is also obvious that a number of points plot outside the triangle defined by the range of possible values. There are a number of possible explanations for this: The scatter of the measurements can be quite high. Another reason might lie in the choice of a wrong mixing law for thermal conductivity. Fig. 7 shows the same data set, but based on an arithmetic law for thermal conductivity:

$$O = \frac{\lambda_f - \lambda}{\rho - \rho_f} . \qquad (9)$$

This way, the data points plot much better within the triangle defined by the mineral end members. However, the data is now less consistent with the compositional information which suggests a higher quartz content.

**Fig. 6** M-O plot of the Bunter data set using a geometric mixing law for thermal conductivity. Blue points are computed from measurements using equations 6 and 8. Red circles mark the (M,O) points expected from the volumetric composition derived from XRD-analysis. Grey lines represent a ternary triangle, spanning the volumetric space between the end member points assuming a quartz-orthoclase-illite mixture. Grid lines are plotted at 20 % intervals.

**Fig. 7** M-O plot of the Bunter data set using an arithmetic mixing law. Blue points are computed from measurements using equations 6 and 9. Red circles mark the (M, O) points expected from the volumetric composition derived from XRD-analysis. Grey lines represent a ternary triangle, spanning the volumetric space between the end member points assuming a quartz-orthoclase-illite mixture. Grid lines are plotted at 20 % intervals.

An advantage of this model over the univariate statistical analysis is that it can be applied to wireline logs run in boreholes. These comprise an additional important source of information. It serves two purposes: (1) The variability of the in situ petrophysical properties can be assessed better from wireline data compared to core data which might be subject to preferential sampling; (2) the large number of boreholes allows a better spatial characterization of changes in facies and corresponding changes in petrophysical properties.

Readings of wireline logs respond to the composition of the probed rock, its structure, and environmental conditions. For the analysis of borehole geophysical data with respect to quantifying rock composition the assumption is made that a log reading responds mainly to the composition of the rock. Then, given some appropriate mixing law and using standard procedures (e.g. Doveton 1979; Hartmann et al. 2005), the lithologic composition can be computed.

For the Upper Jurassic formation, thermal conductivity can be inferred from two geophysical logs which respond to porosity and shale volume, such as, for instance, acoustic slowness (DT) and natural gamma radiation (GR). To analyse logs from the lower Triassic, one additional log is needed. Bulk density may be used, for instance. Special care has to be taken because potassium contained in feldspars might influence the gamma ray log.

A disadvantage of cross-plot methods is that they lack of any measure of uncertainty. The problem is usually posed in such a way, that it is exactly determined. Uncertainties can be visually estimated by the coherence of the cross-plots but a strictly quantitative measure is lacking. This limitation can be overcome by using a full inversion procedure. This is discussed in the next section.

## Advanced data inversion

The following example illustrates the use of an inversion algorithm for the analysis of laboratory measurements performed on a core sample with clearly visible variations in physical properties. The algorithm is described in detail in Hartmann (2007). The inversion uses the mixing laws in order to compute a model data set. The misfit between modelled and measured data is minimised using a Gauss-Newton iterative scheme together with a Bayesian regularisation (e.g. Tarantola 2005). The forward model can be adapted to a particular set of measurements and mixing laws to be used. Usually equations 1 to 3 are used when sonic velocity, density, and thermal conductivity are considered.

This example is particularly suited because of the good control on the quality of measurements and because detailed analyses were easy to perform and can be compared to the actual geology of the rock. The core (Fig. 8, d) was recovered at a depth of about 1300 m in a borehole in the southern German Molasse Basin (Fig. 1 and 2) from the middle Triassic period just above the boundary to the lower Triassic. The lowermost part of the middle Triassic in southern Germany is characterised by massive occurrence of sulphates (anhydrite or gypsum) with a thicknesses of up to 5 m. Thin layers of shaly dolomites are spread throughout this sequence. The structure corresponds to a successive evaporation sequence with temporary decrease of the salt concentration with concurrent enhancement of wave action (Geyer and Gwinner 1991). This structure is reflected in the sample with its dark bands of dolomite embedded in the brighter anhydrite.

**Fig. 8** Measurements taken on a layered anhydrite/dolomite sample. a) Thermal conductivity $\lambda$ (black) and acoustic velocity $v_p$ (grey); b) bulk density $\rho$ measured by gamma-absorption (black) on a core logger (Cl) device and with a powder pycnometer (grey); c) porosity $\phi$ assuming anhydrite (solid black) and dolomite (dashed black) as the matrix compared to pycnometer derived porosity; d) Core photograph showing the layering. Bright bands consist mainly of anhydrite; dark bands are composed mainly of dolomite. Arrows denote downward direction in the borehole

**Table 2** Results of the mineralogical analysis and pycnometer measurements on samples taken from the core. Position X of the plugs along the axis is given in mm. Composition is reported in weight percent. $\rho_{m,c}$ is the matrix density (kg m$^{-3}$), computed from the mineral composition and tabulated mineral densities (see e.g. Wohlenberg, 1982). Pycnometer measurements: Matrix density $\rho_m$ (kg m$^{-3}$) and porosity $\phi$ (%).

To analyse the sample, thermal conductivity, acoustic velocity, and bulk density were measured on the dry sample along the core axis (Fig. 8, a-b). In addition, matrix density and bulk density were determined on three plugs using a pycnometer (Fig. 8, b, c). Porosity is computed along the core using the bulk density measurement (equation 2) assuming pure dolomite ($\rho$ = 2870 kg m$^{-3}$) and pure anhydrite ($\rho$ = 2960 kg m$^{-3}$). In comparison with the pycnometer-derived porosity, Fig. 8, c shows discrepancies suggesting that the sample consists of a mixture of the two minerals. This is confirmed by a mineralogical analysis of two plugs (Table 2). The bright bands are composed mainly of anhydrite whereas the dark bands contain a mixture of both dolomite and anhydrite. The mineral ankerite is chemically and structurally similar to dolomite, with magnesium largely replaced by iron. Thus, ankerite will be added to the volume fraction of dolomite because of its similarity and small volume fraction.

**Fig. 9** M-O plot of the analysed sample. Black dots denote data computed from the measurements on the sample. Gray squares mark expected points for pure minerals. Slowness and density values are taken from Hearst et al. (2000), thermal conductivity from Čermák and Rybach (1982).

Using the same techniques developed in the last section, an M-O plot is constructed for the core scanning data (Fig. 9). The plot shows that the measurements are consistent in a qualitative manner with a mixture composed primarily of anhydrite and dolomite. However, the M-O plot suggests large amounts of dolomite for some measurements, whereas the mineralogical results imply that the grey bands contain only about 50 % dolomite.

There are two possible explanations for these observations that are studied by modifying the inverse model: (1) The particular mixing laws used for sonic velocity and thermal conductivity may be inadequate for low-porosity chemical sediments. To study this uncertainty, a general mixing law that accounts for structural effects is implemented (Korvin 1978), described in detail below. It introduces an additional parameter to deal with sample of varying structure. (2) The uncertainty of mineral thermal conductivities is quite large, in particular for kaolinite. Values for shales and clays are poorly defined and usually assumed around 2.0 W $(\mathrm{m\,K})^{-1}$ (Brigaud and Vasseur 1989). However, Waples and Tirsgaard (2002), for instance, found in their study of clay thermal conductivities a systematic change with depth which they interpreted as due to a increasing orientation of clay minerals with overburden pressure: Thermal conductivity decreased from 2.5 W $(\mathrm{m\,K})^{-1}$ to 1 W $(\mathrm{m\,K})^{-1}$ and the anisotropy factor (the ratio of thermal conductivities parallel over perpendicular to the direction of maximum thermal conductivity) increased from 1 to 2. Both of these effects increase the uncertainty of the position of the pure minerals. To address this, the physical properties of the mineral are incorporated in the inversion with a priori variances assigned.

The inverse model consists of five volume fractions with a priori values and corresponding standard deviations derived from the chemical analysis: Quartz (0.02 ± 0.01), Anhydrite (0.45 ± 1.0), Dolomite (0.45 ± 1.0), Kaolinite (0.05 ± 0.01), and air filling of the pore space (0.03 ± 0.02). Values for physical properties are taken from the literature (Hearst et al. 2000; Čermák and Rybach 1982). The data are assumed to be uncorrelated with standard deviations of the measurements of 0.05 W $(\mathrm{m\,K})^{-1}$ for thermal conductivity, 50 m $\mathrm{s}^{-1}$ for sonic velocity, and 5 kg $\mathrm{m}^{-3}$ for density. A set of different models was run in order to test several hypotheses. Results are summarised in Table 3.

The most simple model (Table 3, model 1) employs the time-average formula for sonic velocity $v_p$ (Wyllie et al 1956), and the geometric mixing law for thermal conductivity $\lambda$. It can be considered a benchmark as it is closest to both the direct transformation shown in the M-O plot (Fig. 9) and results that may be obtained from a more conventional inversion approach. The result (Fig. 10, top) confirms the interpretation of the M-O plot which suggests that the model is inconsistent with the data. No possible volumetric composition within this model can fit the data.

To refine the model, the $t^{\mathrm{th}}$-order mean model by Korvin (1978) is introduced (see also Hartmann 2007). This is a general mixing law yielding an effective value $M$ based on fluid properties $M_f$, matrix properties $M_m$, and a structural parameter $t$.

$$M = \left[\phi\, M_f^{\,t} + (1-\phi)\, M_m^{\,t}\right]^{1/t}, \qquad \text{with } -1 \le t \le 1 \tag{10}$$

The special cases $t$ = -1, 0, +1 correspond to the harmonic, geometric, and arithmetic mixing laws, respectively. To test the hypothesis that a single parameter $t$ can be used to explain the data, two models are designed (Table 3: models 2 and 3), one with a single $t$ and one with two independent $t$-values for $v_p$ and $\lambda$. Values for $t$ in the range of 0.8 - 1.0 yield the best data fit, strongly indicating that standard mixing laws are invalid for well lithified chemical sediments. Model 3 yields slightly different values for $t$ for $v_p$ and $\lambda$, but the difference is insignificant within the error bounds. Both models reduce the misfit. However, the fit is still inconsistent with the data, suggesting that uncertainty about the appropriate mixing law cannot be the only cause.

Therefore, the same set of models is modified to include variable mineral thermal conductivities in the inversion (Table 3: models 4 to 6). Thermal conductivity values of the minerals are assigned a priori standard deviations of 0.1 W $(m\ K)^{-1}$. Only anhydrite, dolomite, and kaolinite are included in the inversion. Using the geometric and Wyllie's average (Wyllie et al. 1956) (model 4) yields an improved but still large misfit of 1.62 W $(m\ K)^{-1}$. In addition, the thermal conductivity value returned for kaolinite is unrealistically low.

The best results are obtained for models 5 and 6 which use the general mixing law both for sonic velocity and thermal conductivity, together with an inversion of the mineral properties. The result for model 6 is shown in Fig. 10 (bottom). The RMS-error for both models is below one, indicating a slight over-fitting of the data because the RMS-error is normalised by the measurement error. A value lower one indicates that noise in the data is fitted. It can be noted that an inverse correlation exists between the value of $t$ and the matrix thermal conductivities for models 4 to 6. The mineral thermal conductivities of anhydrite and dolomite are large but plausible, given the range of values in the literature (Clauser and Huenges 1995; Clauser 2006). The clay mineral value is quite low. It is most strongly affected by the choice of model, and due to the generally low content of kaolinite the least trustworthy.

Models 5 and 6 show that $t$-values larger than 0.6 are required to fit the measured data. Note that a value of 0.5 corresponds to the square root average sometimes used for thermal conductivity (Korvin 1978; Beardsmore and Cull 2001). The geometric mean has been confirmed in many studies to be adequate for the analysis of sedimentary rocks (e.g. Sass et al. 1971; Brigaud et al. 1990; Hartmann et al. 2005). However, it can be reasoned that the structure of the well-lithified chemical sediment studied here is quite different from that of granular sediments. The RMS misfit between modelled and measured data can be estimated to be the predictive error of the method. The RMS values are 0.07 W $(m\ K)^{-1}$, 70 m $s^{-1}$, and 8 kg $m^{-3}$ for thermal conductivity, sonic velocity, and bulk density, respectively. Expressed as percentages of the average measured value, these correspond to relative errors of 1.3 %, 1.2 %, and 0.3 % for thermal conductivity, sonic velocity, and bulk density.

There is no conclusive answer to the question whether the same mixing law might be used for both thermal conductivity and sonic velocity. Both display large $t$ values, the effect apparently being stronger for sonic velocity than for thermal conductivity. However, some ambiguity remains because of the inverse relationship between matrix thermal conductivity and $t$-value noted above. This correlation makes the high $t$-values less reliable. A second point is the large standard deviation of the $t$-value, making inferences about their differences in $t$ difficult.

This example demonstrates that a sophisticated inversion yields a very detailed description and understanding of the petrophysical relationships. Using a Bayesian framework, a-posteriori uncertainties can be specified which are based both on the uncertainty in the input data and on the uncertain information on property values or even mixing laws. In conclusion, this type of analysis provides the best possible characterisation of a given petrophysical data set. Clearly, it also the most complex method and is usually performed only on selected subsets of the data.

**Fig. 10** Results of the inversion of high resolution measurements for the anhydrite/dolomite sample. Index *d* refers to measured data; index *m* refers to modelled data. Top: Results for model 1. Bottom: Results for model 6. a) – c) Measured data and computed values for thermal conductivity, acoustic velocity, and bulk density; d) Mineral composition of the sample.

**Table 3** Summary of inverted mineral compositions and physical properties for different petrophysical models. The "Data" column shows measured values ($v_p$ in m $s^{-1}$, $\lambda$ in W (m K)$^{-1}$; $\rho_b$ in kg $m^{-3}$). These are compared to inverted parameters for models 1 to 6. See text for a more detailed discussion.

# Conclusion

From the discussion of the methods it is clear that all of them have certain advantages but also suffer from specific disadvantages. Univariate descriptive statistics have the major advantage of being easy to understand and use. This is particularly important when results are used in environments with restricted time and resources. The major drawback is that for some lithologies the uncertainty of the derived estimates may be quite large. Also, information on the local geology may be difficult to use, since there is no formal way to include this type of information in univariate descriptive statistics. This can be somewhat allowed for by dividing the data set into sub-units of similar lithology. However, stratigraphic units cannot always be identified with lithologic units. For the studied stratigraphic units, uncertainty of thermal conductivity predictions ranges from 0.8 to 0.1 W (m K)$^{-1}$. The lower uncertainties can be achieved for the Upper Jurassic data sets, where lithology is relatively uniform and porosity is not a dominant factor. On the other hand the Lower Triassic data have a wider range of composition and porosity values, leading to much larger error margins.

Cross-plot methods partly overcome this limitation since they can be used to display the petrophysical relationships. In ternary diagrams, petrophysical properties can be read directly from the chart if the lithology can be estimated. Here, the applicability of the geometric mixing law was verified by comparing petrophysical measurements and mineralogical data in the same ternary diagram. However, they only yield a single estimate without uncertainty, a significant drawback when simulating technical or natural systems. This can be circumvented to some extent by estimating the most likely ranges although this is not an uncertainty estimate in the strict sense.

Inversion methods are capable of both: providing strict uncertainty estimates plus and incorporating multiple sources of information. Within a Bayesian framework, data and their covariance as well as a priori information are used to estimate the maximum a-posteriori estimate jointly with its covariance. The example presented illustrates that very detailed analyses can be applied including several types of measurements. Predictive errors in the low percentage range are possible for a detailed analysis, restricted mostly by the accuracy of measurement equipment in the laboratory. Mineral thermal conductivities can be modelled to an accuracy of 0.2 W (m K)$^{-1}$. Of course, this type of application is rather complex and possibly restricted to specialised studies.

In summary, a combined approach seems to be most reasonable, providing the analyst with a choice for the most suitable model for his specific data set. Specifically, the methods discussed provide a means to introduce data uncertainty into the analysis of both studies of basin-scale heat and fluid flow and prognostic design calculations of technical installations for geothermal energy use.

**Acknowledgments**: This study was funded by the German Federal Ministry for the Environment, Nature Conversation, and Nuclear Safety under grant No. 0329985 to RWTH Aachen University. Lothar Ahrensmeier and Dirk Breuer performed most of the laboratory measurements. Rushana Valiyeva and an anonymous reviewer provided helpful comments that improved the quality of the manuscript.

| | Upper Bunter | | | | Middle Bunter | | | |
|---|---|---|---|---|---|---|---|---|
| | 25% | 50% | 75% | N | 25% | 50% | 75% | N |
| This study | 3.74 | 3.84 | 3.94 | 133 | 4.16 | 4.49 | 4.91 | 748 |
| Kleiner (2003) | 4.04 | 4.25 | 4.61 | 30 | 4.98 | 5.23 | 5.43 | 92 |

| | X | Quartz | Anhydrite | Dolomite | Kaolinite | Ankerite | $\rho_{m,c}$ | $\rho_m$ | $\phi$ |
|---|---|---|---|---|---|---|---|---|---|
| Plug 1 | 60 | 0.52 | 89.84 | 2.44 | 5.62 | 1.58 | 2942 | 2944 | 2.7 |
| Plug 2 | 175 | | | | | | | 2866 | 2.0 |
| Plug 3 | 215 | 3.09 | 43.92 | 48.83 | 4.16 | | 2896 | 2895 | 3.3 |

| | | Data | Model 1 | Model 2 | Model 3 | Model 4 | Model 5 | Model 6 |
|---|---|---|---|---|---|---|---|---|
| $v_p$ model | | | Wyllie | joint $t^{th}$ order mean | single $t^{th}$ order mean | Wyllie | joint $t^{th}$ order mean | single $t^{th}$ order mean |
| $\lambda$ model | | | geometric | | single $t^{th}$ order mean | geometric | | single $t^{th}$ order mean |
| normalised *RMS* error | | | 2.29 | 1.96 | 1.95 | 1.62 | 0.85 | 0.86 |
| $t_\lambda$ | | | | 0.99 ± 0.38 | 0.97 ± 0.38 | | 0.90 ± 0.37 | 0.62 ± 0.36 |
| $t_{vp}$ | | | | | 0.80 ± 0.36 | | | 0.89 ± 0.38 |
| Plug 1 volumes [%] | $V_{anhydrite}$ | 89.8 | 86.3 | 88.5 | 88.5 | 85.9 | 87.2 | 87.1 |
| | $V_{dolomite}$ | 4.0 | 7.1 | 4.5 | 4.7 | 6.7 | 3.8 | 3.8 |
| | $V_{kaolinite}$ | 5.6 | 4.6 | 2.8 | 2.8 | 5.4 | 4.8 | 4.9 |
| | $\phi$ | 2.7 | 0.3 | 2.4 | 2.4 | 0.3 | 2.6 | 2.6 |
| Plug 3 volumes [%] | $V_{anhydrite}$ | 43.9 | 43.8 | 44.1 | 44.0 | 42.9 | 43.0 | 43.1 |
| | $V_{dolomite}$ | 48.8 | 48.8 | 48.7 | 48.8 | 48.3 | 48.2 | 48.1 |
| | $V_{kaolinite}$ | 4.2 | 5.6 | 2.8 | 2.8 | 7.0 | 4.4 | 4.4 |
| | $\phi$ | 3.3 | 0.1 | 2.7 | 2.7 | 0.1 | 2.8 | 2.8 |
| Mean volumes [%] | $V_{anhydrite}$ | | 39.0 | 64.3 | 62.8 | 16.9 | 58.0 | 58.5 |
| | $V_{dolomite}$ | | 55.6 | 29.0 | 30.7 | 76.0 | 33.0 | 32.4 |
| | $V_{kaolinite}$ | | 3.1 | 2.1 | 2.1 | 4.7 | 4.7 | 4.7 |
| | $\phi$ | | 0.5 | 2.8 | 2.7 | 0.7 | 2.7 | 2.7 |
| Mean sample values | $\lambda$ | 4.99 | 4.84 | 4.77 | 4.77 | 5.04 | 4.98 | 4.99 |
| | $v_p$ | 5864 | 5869 | 5875 | 5869 | 5802 | 5853 | 5843 |
| | $\rho_b$ | 2826 | 2875 | 2835 | 2836 | 2846 | 2827 | 2827 |
| Estimated mineral values | $\lambda_{anhydrite}$ | | | | | 6.20 ± 0.22 | 6.03 ± 0.17 | 6.12 ± 0.20 |
| | $\lambda_{dolomite}$ | | | | | 5.97 ± 0.20 | 5.31 ± 0.14 | 5.38 ± 0.20 |
| | $\lambda_{kaolinite}$ | | | | | 0.29 ± 0.20 | 1.21 ± 0.28 | 1.23 ± 0.28 |

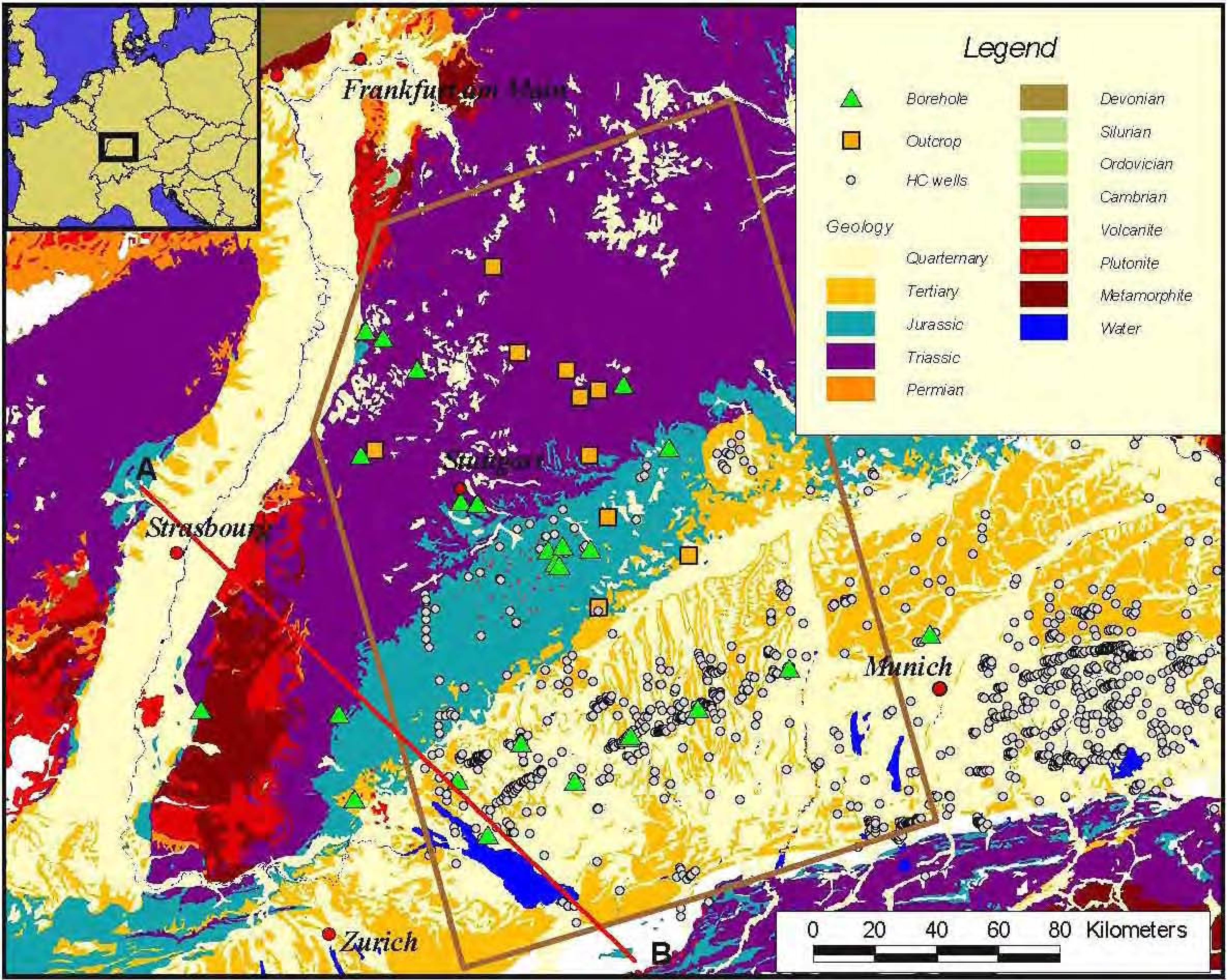

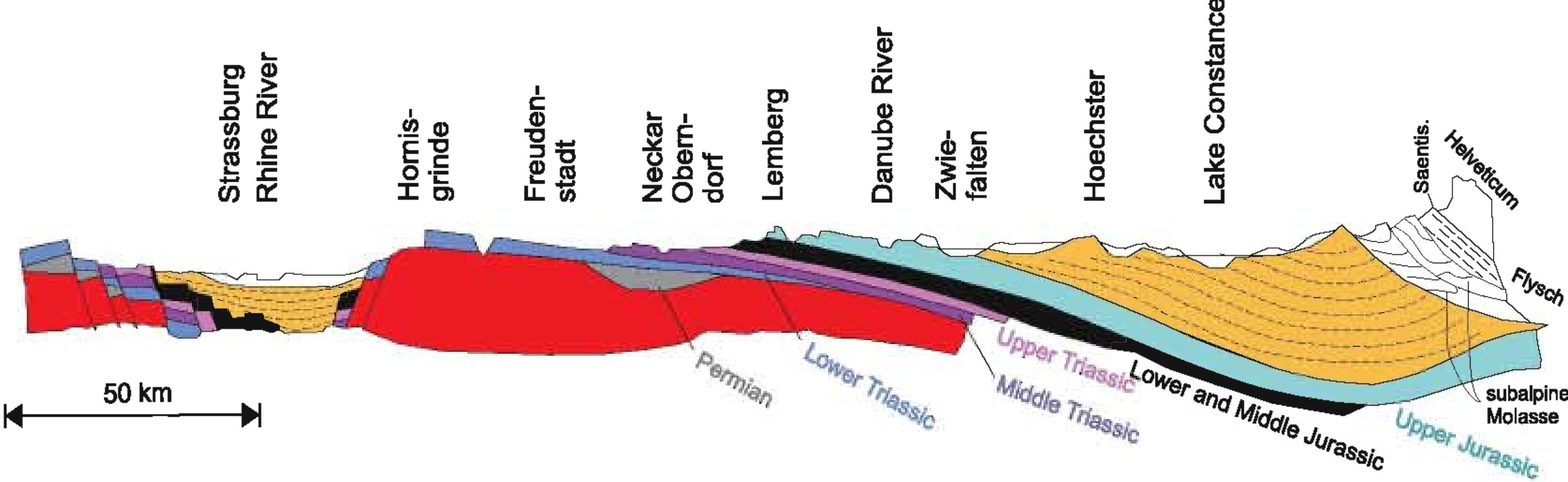

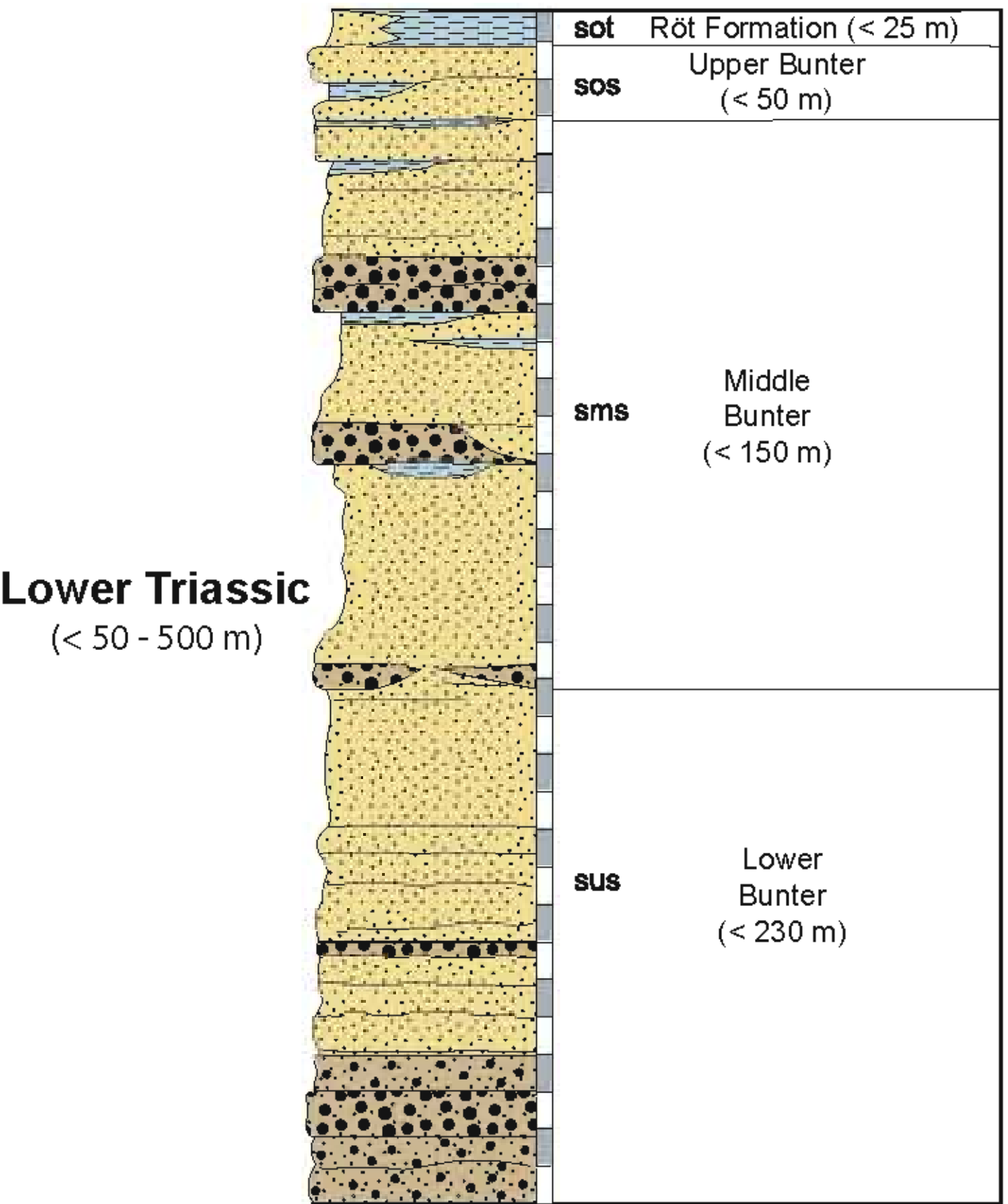

Lower Triassic
(< 50 - 500 m)
sot Röt Formation (< 25 m)
sos Upper Bunter (< 50 m)
sms Middle Bunter (< 150 m)
sus Lower Bunter (< 230 m)

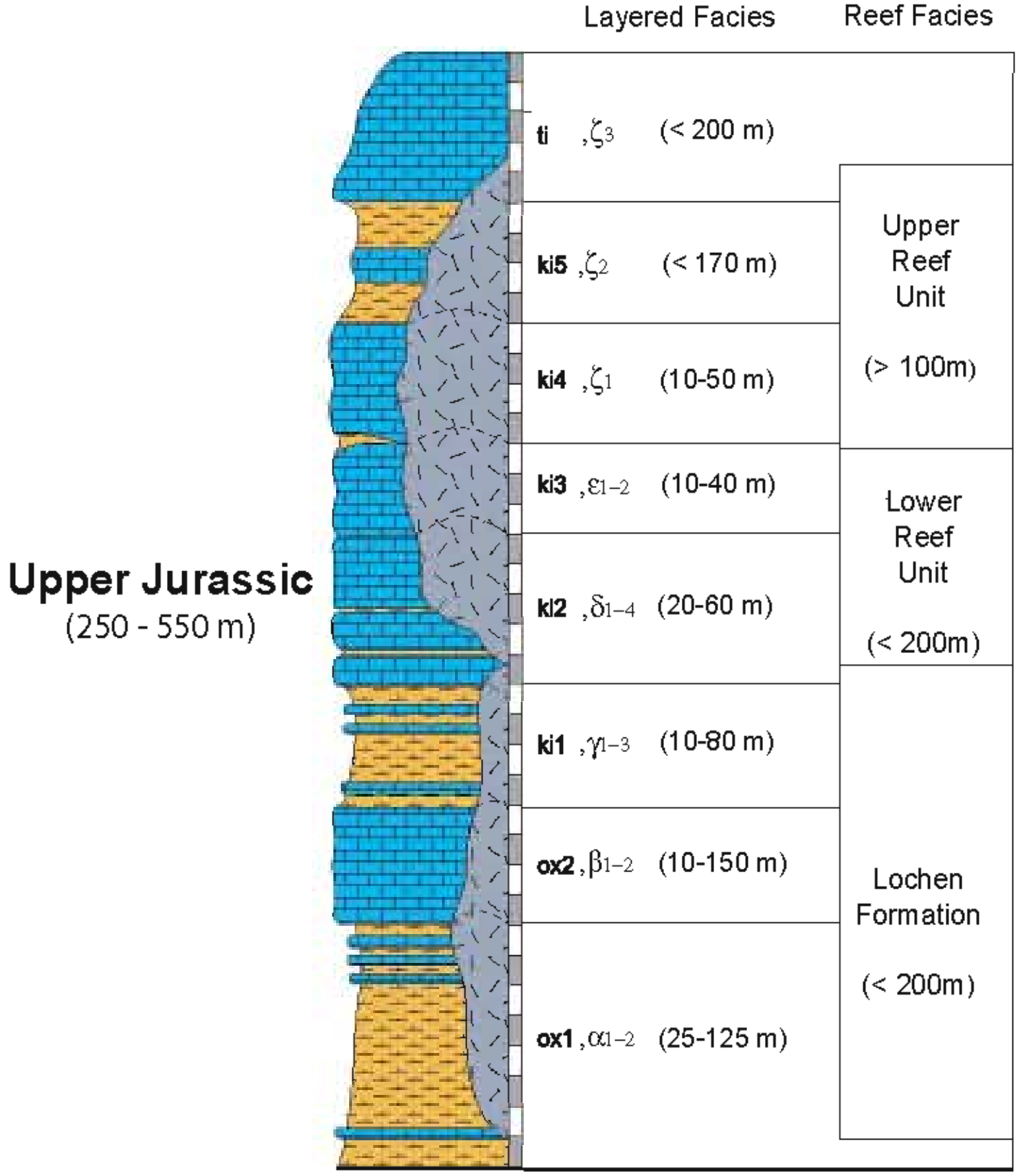

Upper Jurassic
(250 - 550 m)
Layered Facies
Reef Facies
ti ,ζ3 (< 200 m)
ki5 ,ζ2 (< 170 m)
ki4 ,ζ1 (10-50 m)
ki3 ,ε1–2 (10-40 m)
ki2 ,δ1–4 (20-60 m)
ki1 ,γ1–3 (10-80 m)
ox2 ,β1–2 (10-150 m)
ox1 ,α1–2 (25-125 m)
Upper Reef Unit (> 100m)
Lower Reef Unit (< 200m)
Lochen Formation (< 200m)

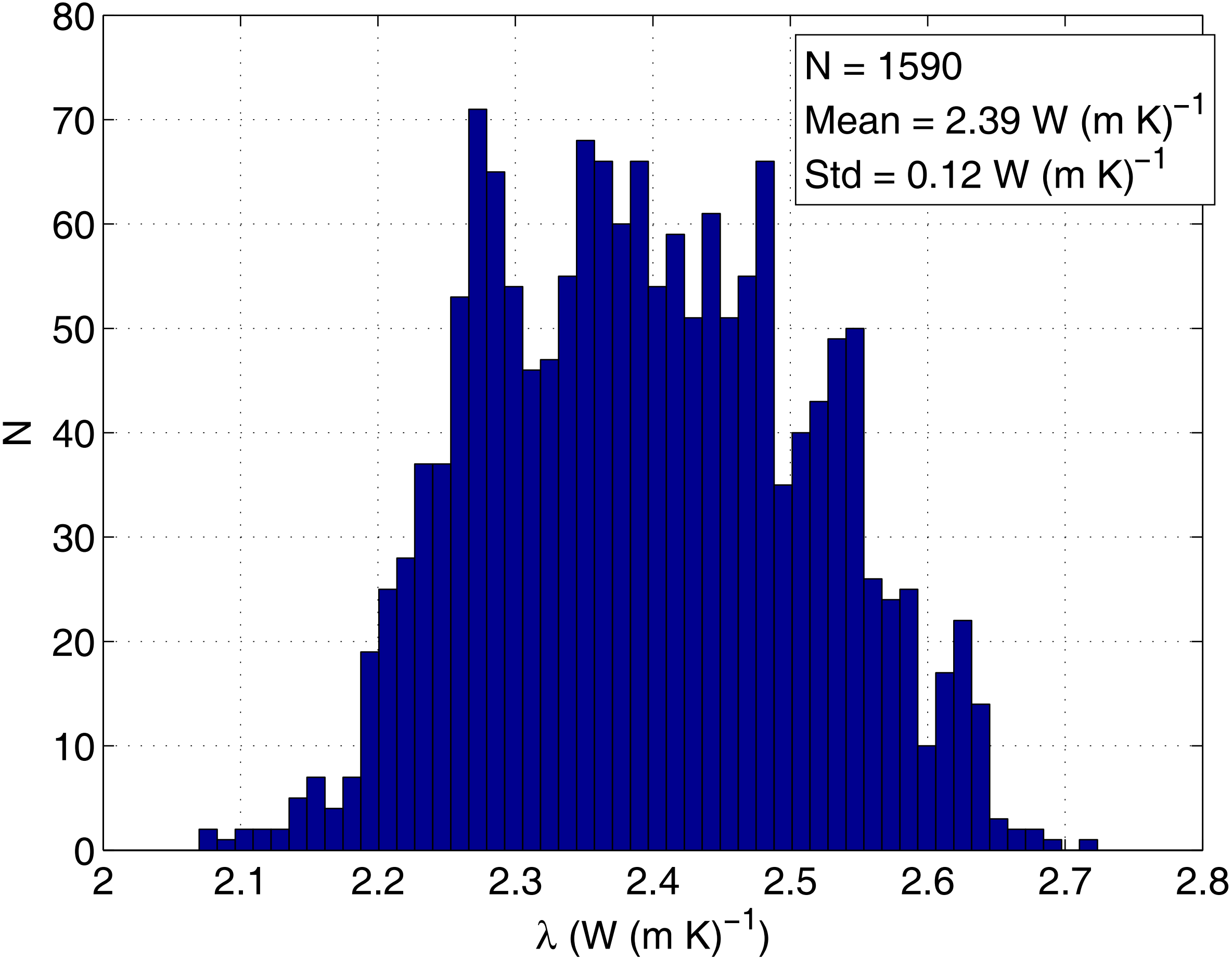

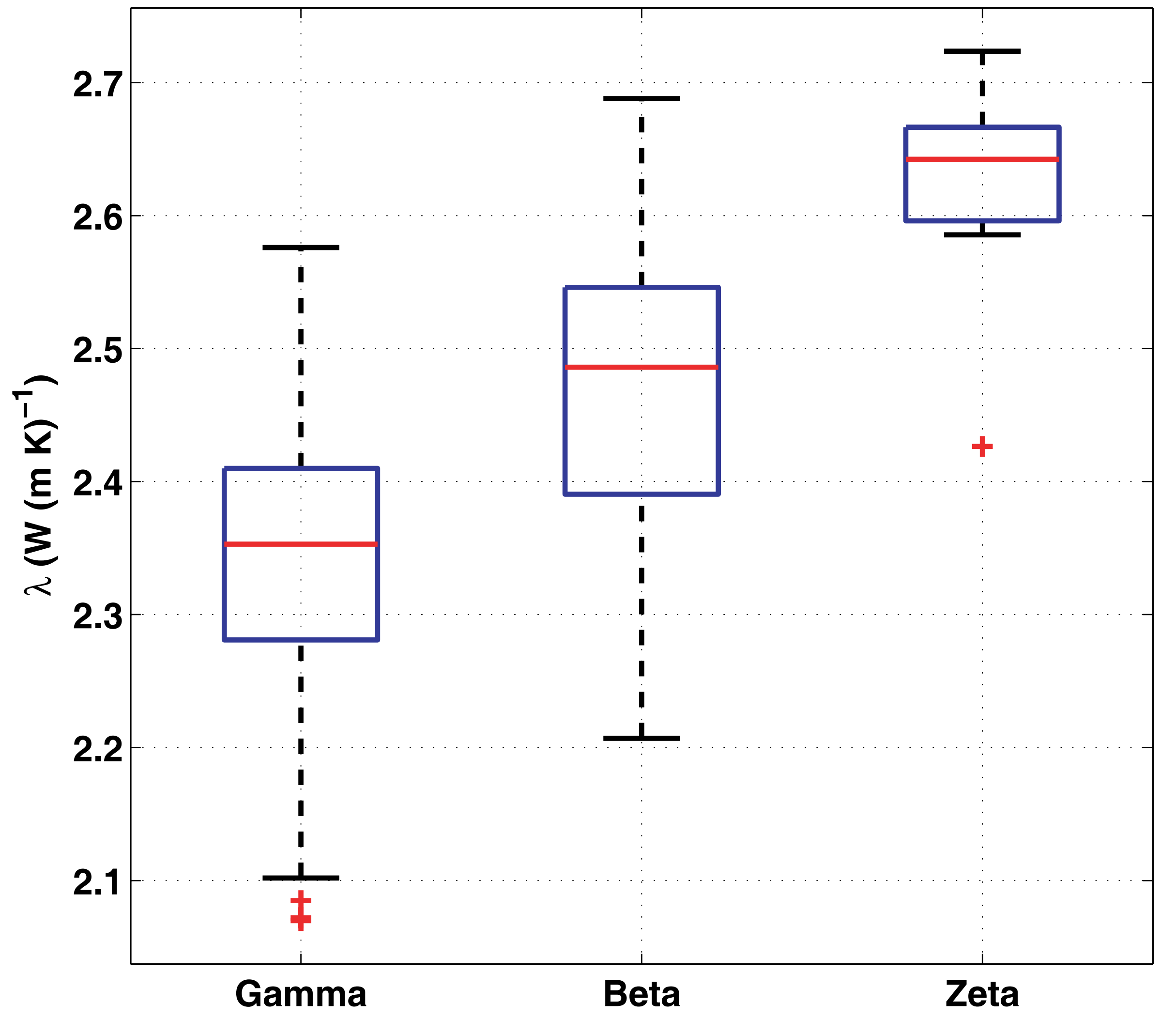

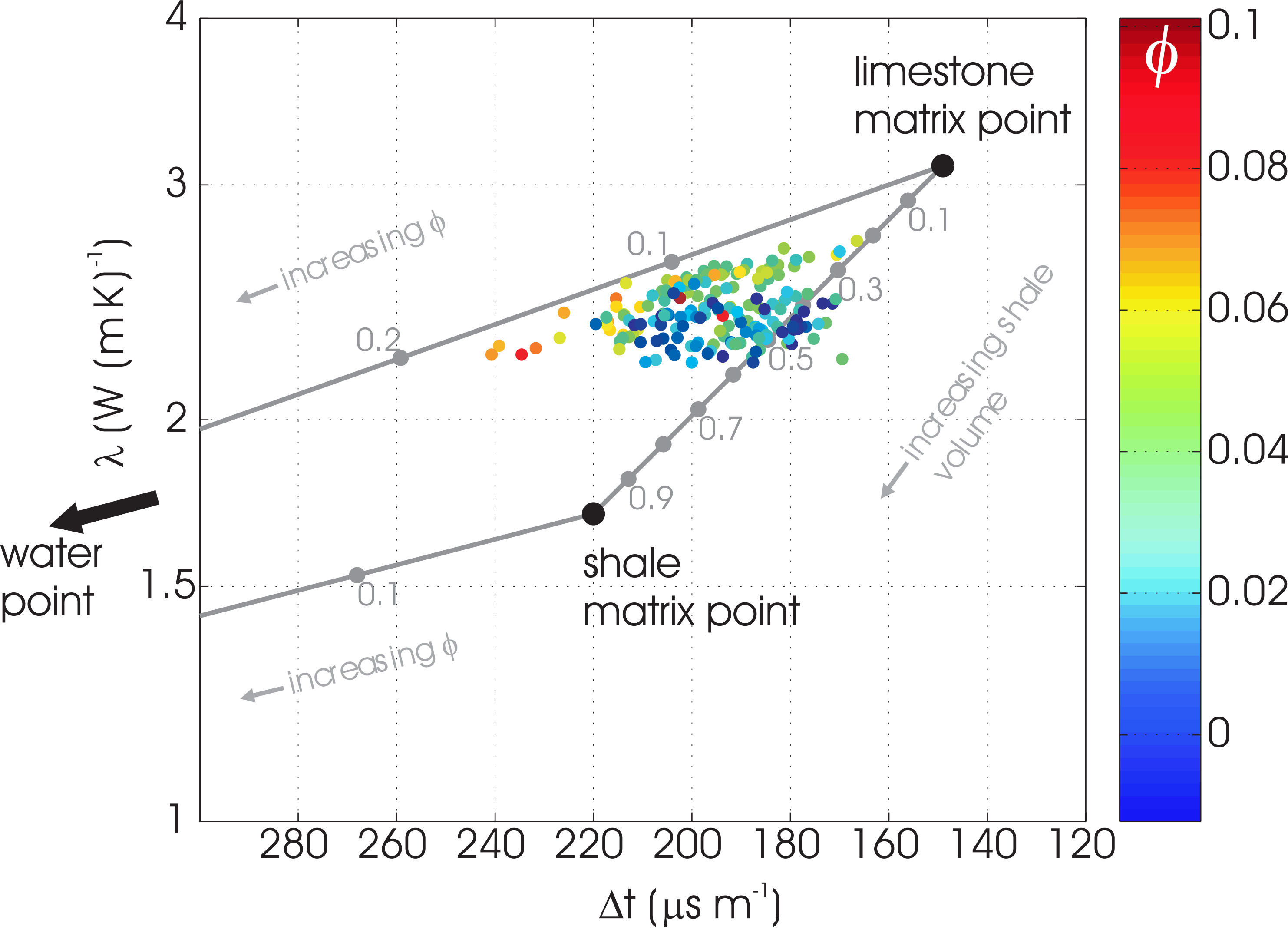

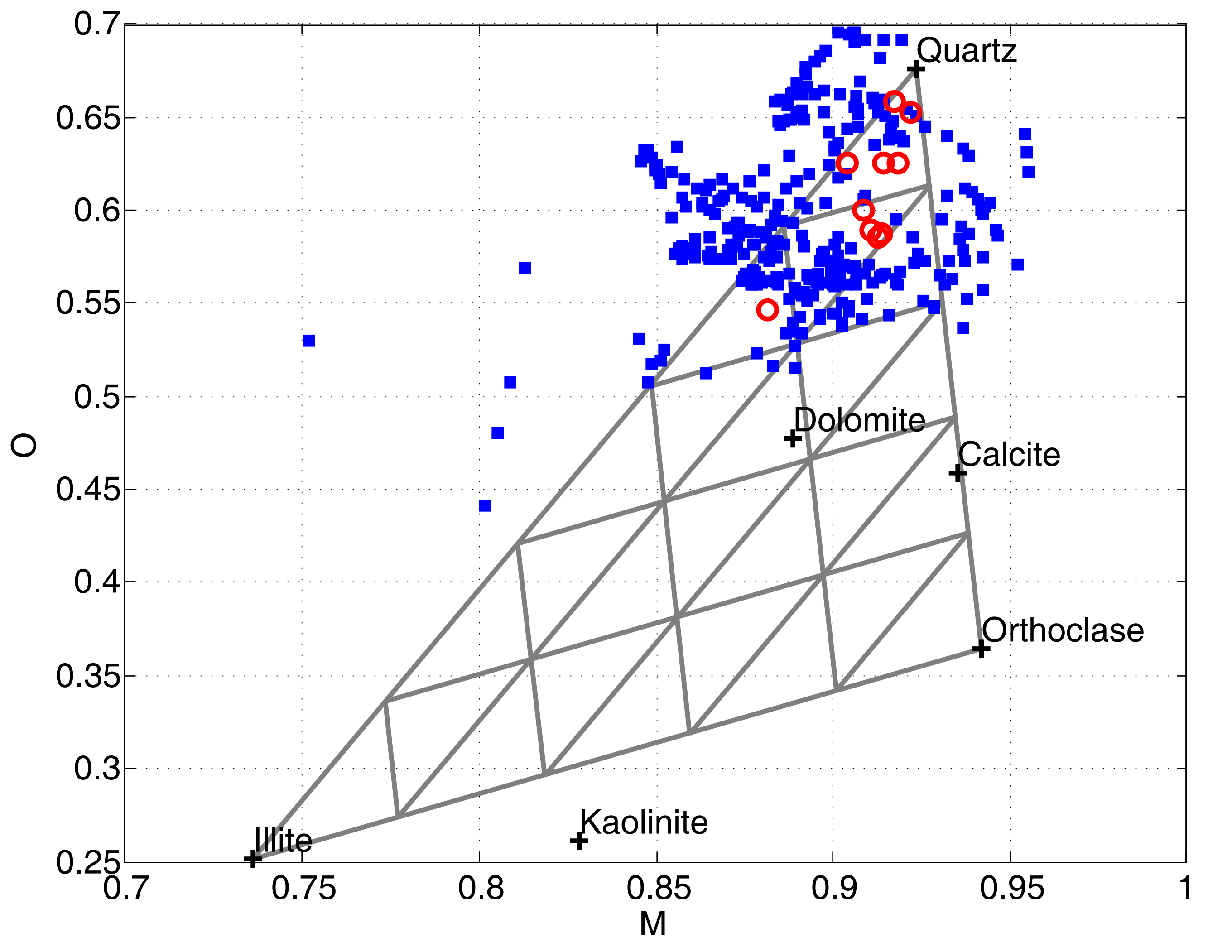

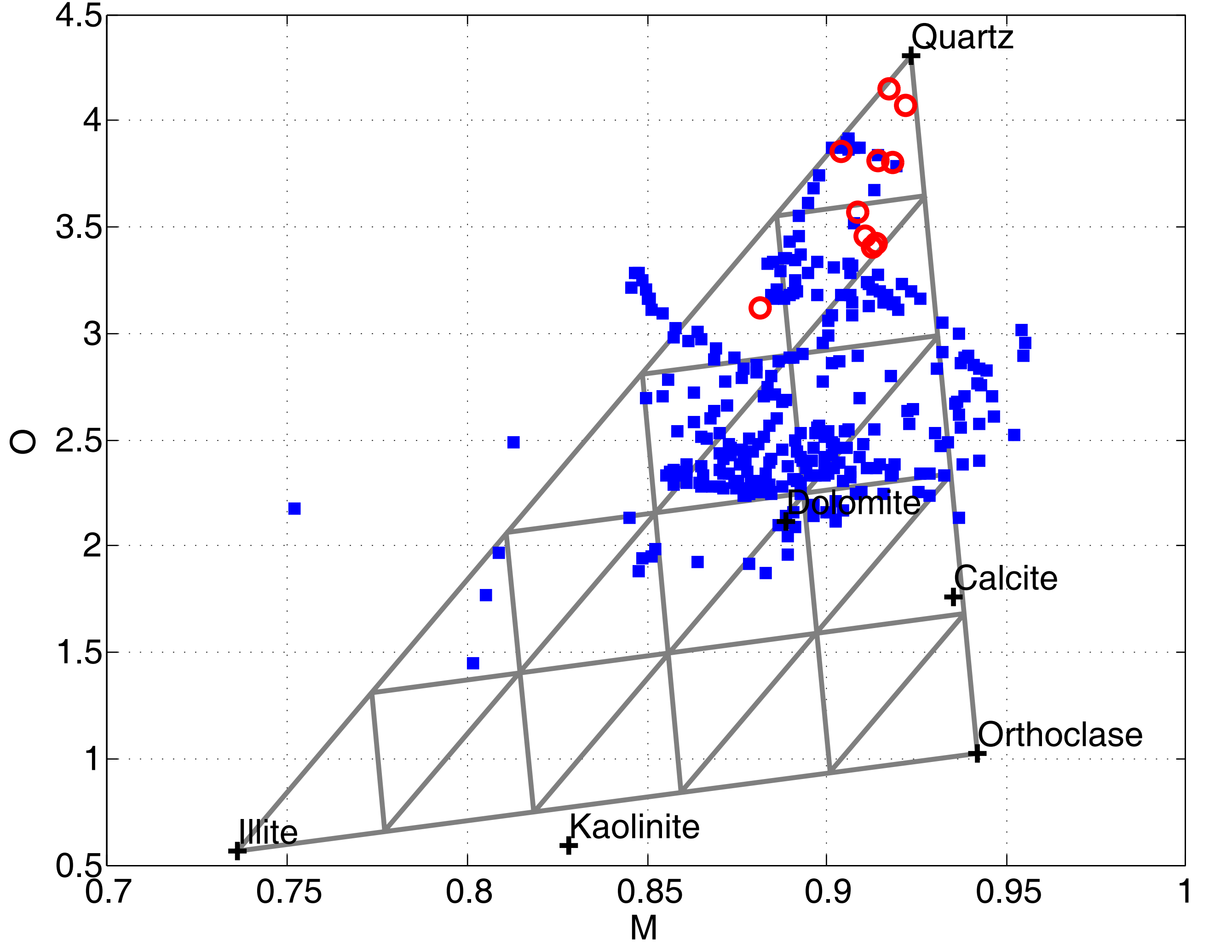

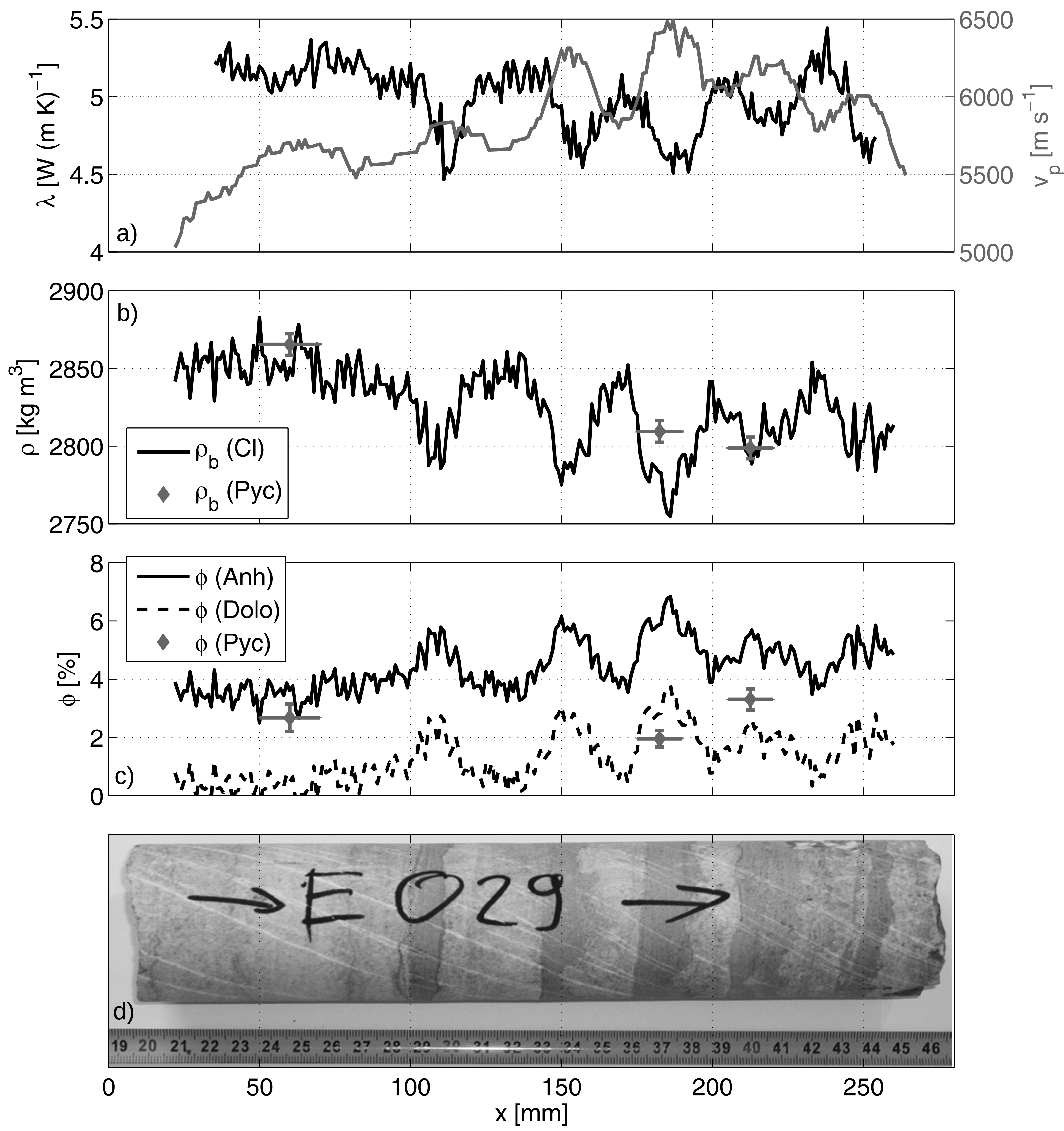

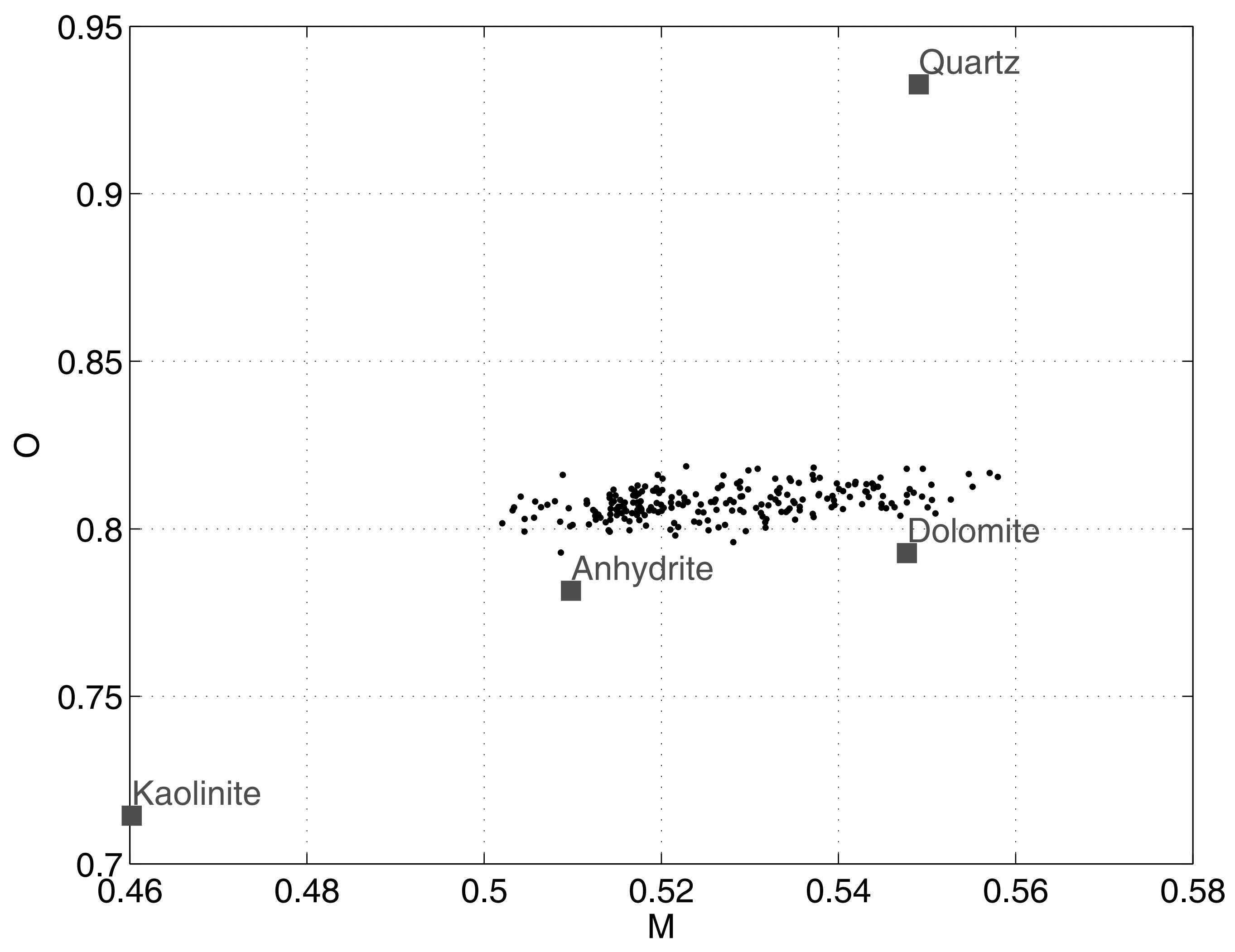

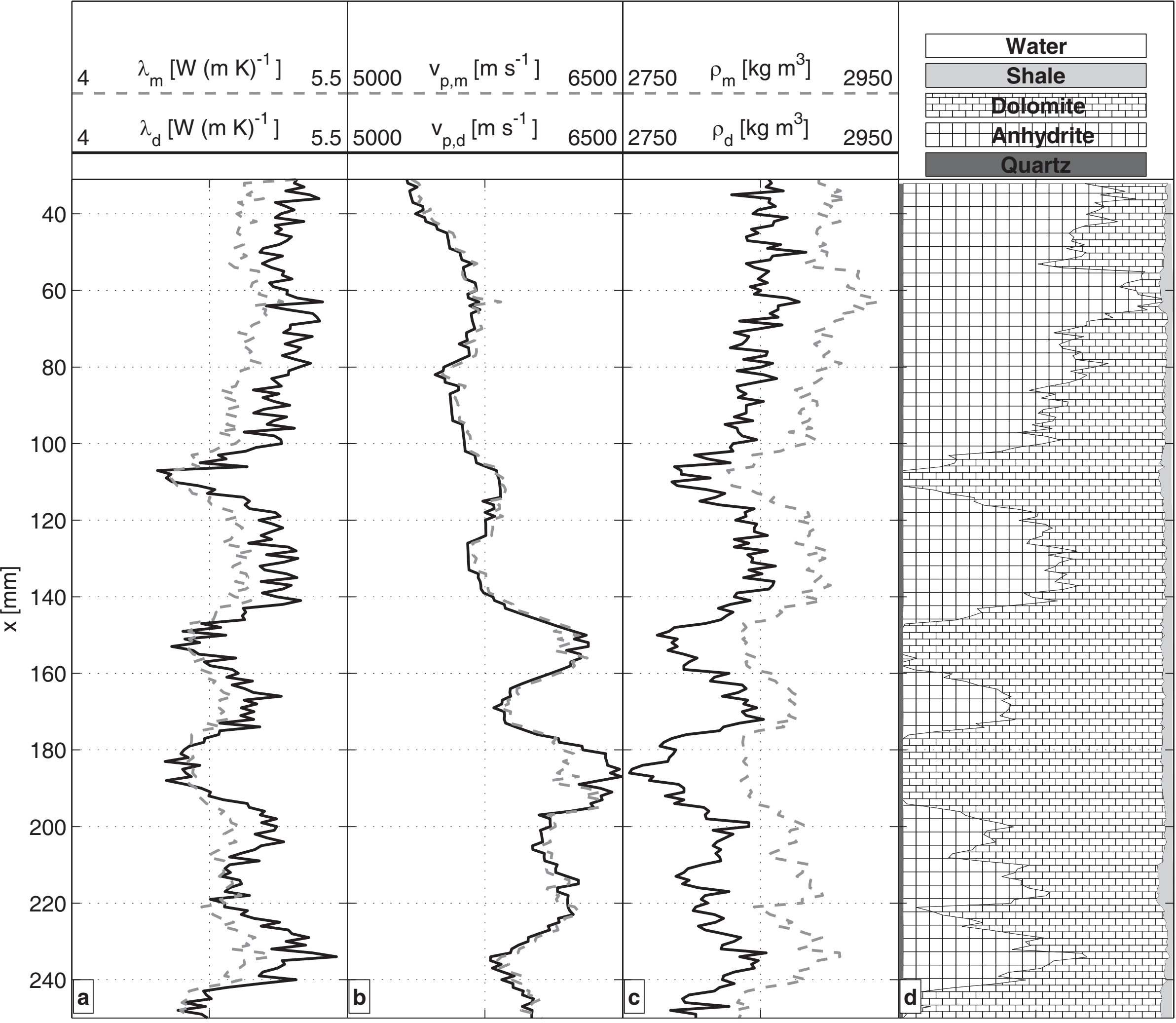
4 $\lambda_m$ [W (m K)$^{-1}$] 5.5
4 $\lambda_d$ [W (m K)$^{-1}$] 5.5
5000 $v_{p,m}$ [m s$^{-1}$] 6500
5000 $v_{p,d}$ [m s$^{-1}$] 6500
2750 $\rho_m$ [kg m$^3$] 2950
2750 $\rho_d$ [kg m$^3$] 2950
Water
Shale
Dolomite
Anhydrite
Quartz
x [mm]
40
60
80
100
120
140
160
180
200
220
240
a
b
c
d

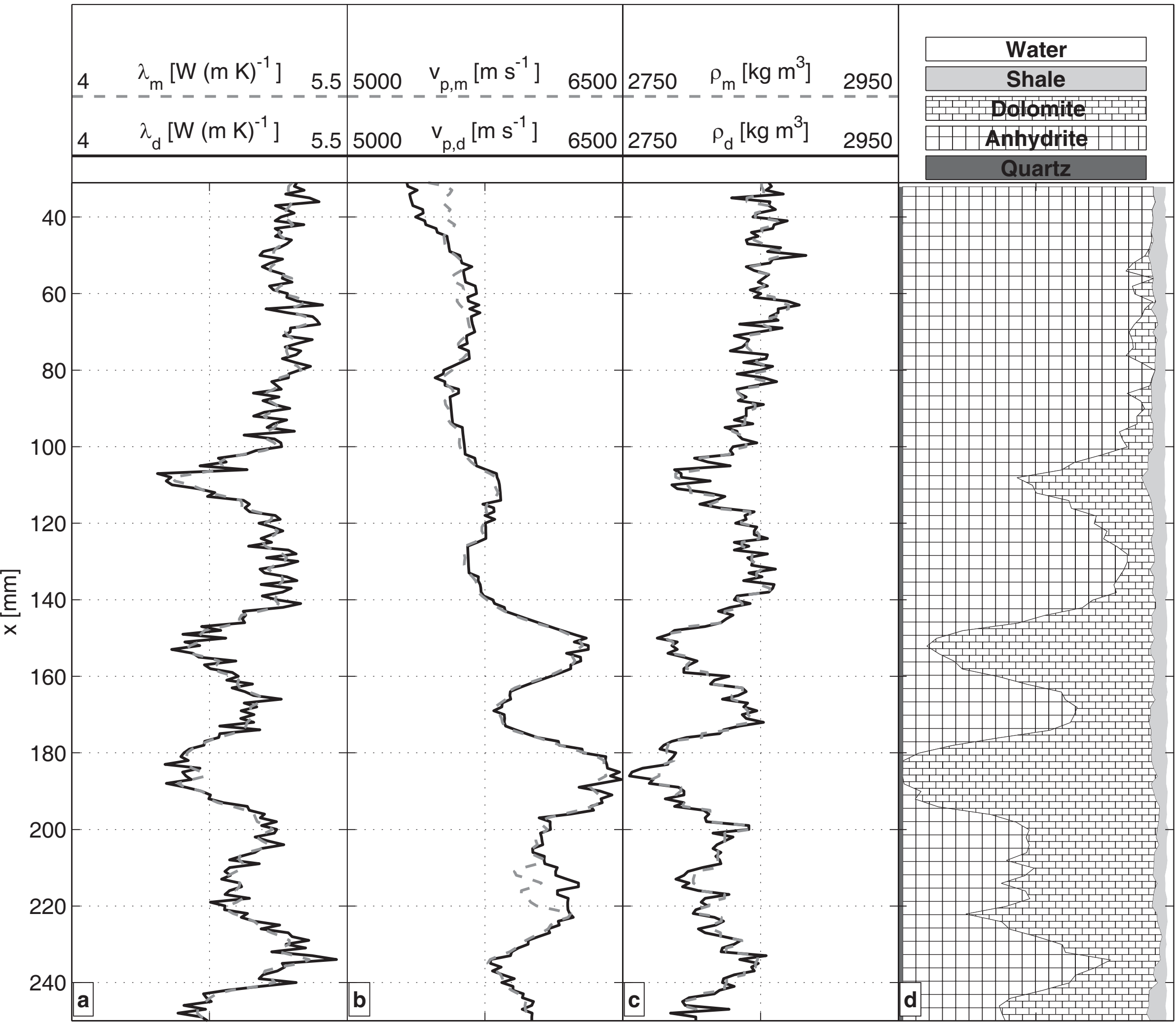
4 $\lambda_m$ [W (m K)$^{-1}$] 5.5
4 $\lambda_d$ [W (m K)$^{-1}$] 5.5
5000 $v_{p,m}$ [m s$^{-1}$] 6500
5000 $v_{p,d}$ [m s$^{-1}$] 6500
2750 $\rho_m$ [kg m$^3$] 2950
2750 $\rho_d$ [kg m$^3$] 2950
Water
Shale
Dolomite
Anhydrite
Quartz
x [mm]
40
60
80
100
120
140
160
180
200
220
240
a
b
c
d